\documentclass[lettersize,journal]{IEEEtran}
\usepackage{amsmath,amsfonts}
\usepackage{algorithmic}
\usepackage{array}
\usepackage{textcomp}
\usepackage{stfloats}
\usepackage{url}
\usepackage{verbatim}
\usepackage{graphicx}
\usepackage{cite}
\usepackage[utf8]{inputenc} 
\usepackage[T1]{fontenc}    
\usepackage{hyperref}       
\usepackage{url}            
\usepackage{booktabs}       
\usepackage{amsfonts}       
\usepackage{nicefrac}       
\usepackage{microtype}      
\usepackage{xcolor}         

\usepackage{graphicx}
\usepackage{textcomp}
\usepackage{xcolor}
\usepackage{amssymb}

\usepackage{booktabs} 
\usepackage{float}
\usepackage{multirow}
\usepackage{indentfirst}
\usepackage{multirow}
\usepackage{amsmath}
\usepackage{graphicx}
\usepackage[]{graphicx,indentfirst}
\usepackage{arydshln}
\usepackage{textcomp}
\usepackage{multicol}
\usepackage{bm}
\usepackage{url}
\usepackage{amsfonts}
\usepackage{mathrsfs}

\usepackage{algorithmic}
\usepackage[ruled,linesnumbered]{algorithm2e}

\usepackage{ulem}
\usepackage{cite}
\usepackage{amsmath,amssymb,amsfonts}
\usepackage{algorithmic}
\usepackage{graphicx}
\usepackage{textcomp}
\usepackage{xcolor}
\usepackage[mathscr]{eucal}
\usepackage{arydshln}
\usepackage{orcidlink}
\usepackage{booktabs}
\usepackage{subfigure}
\usepackage{threeparttable}

\usepackage{mhchem}
\begin{document}

\title{Self-Supervised Learning for Glass Composition Screening}

\author{\IEEEauthorblockN{Meijing Chen\IEEEauthorrefmark{2}\IEEEauthorrefmark{3} \textsuperscript{\orcidlink{0009-0003-5900-1067}} , Bin Liu\IEEEauthorrefmark{4}~\textsuperscript{\orcidlink{0000-0002-1011-2909}} , Ying Liu\IEEEauthorrefmark{2} ,Tianrui Li\IEEEauthorrefmark{4}~\textsuperscript{\orcidlink{0000-0001-7780-104X}} \thanks{ Corresponding authors: Bin Liu, Ying Liu, Tianrui Li.}
		 \thanks{This research was supported by the National Natural Science Foundation of China (Nos. U2468207, 62176221, 61572407,\allowbreak 52472333), Fundamental Research Funds for the Central Universities, China (No.
		 	2682025CX082), Sichuan Science and Technology Program (Nos. \allowbreak2024NSFTD0036, 2024ZHCG0166), Science and Technology R\&D Program of China Railway Group Limited (2023-Major-23), Scientific and Technological Research Program of Chongqing Municipal Education Commission ( No.KJQN202501371), and Fundamental Research Funds for the Central Universities (No.2682025CX082). \\
		 	1359-6454/© 2025 Acta Materialia Inc. Published by Elsevier Inc. All rights are reserved, including those for text and data mining, AI training, and similar
		 	technologies. }
} \\
\IEEEauthorblockA{\IEEEauthorrefmark{2}\textit{College of Materials Science and Engineering, Sichuan University, Chendu, China } \\
\IEEEauthorblockA{\IEEEauthorrefmark{4}\textit{School of Computing and Artificial Intelligence, Southwest Jiaotong University, Chendu, China}}\\
\IEEEauthorblockA{\IEEEauthorrefmark{3}\textit{Research Institute for New Materials Technology, Chongqing University of Arts and Sciences, Chongqing, China}}
	}
}

\markboth{Acta Materialia, Volume 301, 2025, 121509,
 \url{https://doi.org/10.1016/j.actamat.2025.121509}}
{Shell \MakeLowercase{\textit{et al.}}: A Sample Article Using IEEEtran.cls for IEEE Journals}


\maketitle

\begin{abstract}
Glass composition screening is essential for advancing new glass materials, yet the inherent complexity of multicomponent systems presents significant challenges. Current supervised learning methods for this task rely heavily on large amounts of high-quality data and are prone to overfitting on noisy samples, which limits their generalization ability. In this work, we propose a novel self-supervised learning framework designed specifically for screening glass compositions within pre-defined glass transition temperature ranges. We reformulate the screening task as a classification problem, aiming to predict whether the  glass transition temperature of a given composition falls within a target interval. To improvethe model’s robustness to noise, we introduce an innovative data augmentation strategy grounded in asymptotic theory. Additionally, we present \textit{DeepGlassNet}, a dedicated network architecture developed to capture and analyze the complex interactions among constituent elements in glass compositions. Experimental results demonstrate that \textit{DeepGlassNet} achieves superior screening accuracy compared to traditional methods and exhibits strong adaptability to other composition-related screening tasks. This study not only provides an efficient methodology for designing multicomponent glasses but also establishes a foundation for applying self-supervised learning in material discovery. Code and data are available at: \url{https://github.com/liubin06/DeepGlassNet}
\end{abstract}

\section{Introduction}

\IEEEPARstart{G}{lass} has extensive utilities in various fields such as construction materials~\cite{Debije:AEM:2011,Yao:Science:2017,Zhi:2018:RSER}, electronic devices~\cite{Eggleton:2011:NP,Sun:2014:EES,Nie:2024:NC}, and optical instruments~\cite{Arbabi:2014:nc,Xu:2010:NC,Papadopoulos:2019:AM},nuclear waste immobilization~\cite{Icenhower:2015:GECA,Seng:2012:JHM,Xu:2021:AMI}. 
The glass transition temperature (T$_g$), which lies within the range between an equilibrium liquid and a frozen solid, is a vital indicator of glass properties and defines the glassy state ~\cite{Schoenholz:2015:NP,Song:2022:SA,G:1998:JECS}. At this temperature, the structural relaxation time of the glass becomes similar to the experimental observation time, leading to significant changes in physical properties such as atomic structure~\cite{Lin:2021:npj,Schoenholz:2016:PANS}, coefficient of thermal expansion~\cite{Lunkenheimer:2022:NP,F:2004:JECS}, and heat capacity~\cite{Wondraczek:2022:AM,Loretz:2024:JNCS}. 

Currently, the development of new multicomponent glasses with desired T$_g$ is based primarily on empirical methods and the “Edison style” trial and error approach, which are both time-consuming and costly~\cite{Ellison:2003:NM,Liu:2019:JNCS,Lu:2023:JACS}. This challenge is particularly pronounced for oxide glasses, which can be composed of over half of the elements in the periodic table. Consequently, the number of possible compositional combinations far exceeds the capacity of any laboratory to prepare and test them all~\cite{Cassar:2021:CI,Liu:2024:MTC}. Traditional materials prediction methods, such as density functional theory~\cite{Yash:1991:PR,Mandal:2014:NC,Jain:2016:NRM}, molecular dynamics simulations~\cite{Du:2010:JACS,chri:2021：PRM,Tilocca:2011:JMC}, topological constraint theory~\cite{Langer:2014:RPIP,Gin:2020:npjMD,Smed:2010:PRL}, and ab initio methods~\cite{Gaillac:2020:CM,Christie:2014:B}, are generally limited to simple material systems and struggle to handle the complexity of multi-component glass systems~\cite{Alcobaça:2020:Acta}. Machine learning, recognized as a leading research area for the 2024 Nobel Prize in Physics, has demonstrated immense potential in addressing large-scale and complex material systems~\cite{fan:2022:am, Yang:2024:Acta, Liu:2023:Acta, Wu:2023:Acta,Wu:2024:AM,Carvalho:2024:AMI,fan:2022:PRM}, making it an effective approach for developing glasses with specific properties~\cite{Li:2024:npj,Wu:2023:AM}. 
In the past few years, machine learning has found applications in predicting the glass transition temperature (T$_g$)~\cite{Wu:2022:Acta}, glass structures~\cite{Bødker:2022:npj}, glass hardness~\cite{Garg:2023:JACS}, and the coefficient of thermal expansion (CTE) of glass~\cite{Mastelini:2022:Acta}, as well as establishing correlation between glass structures and properties~\cite{ fan:2020:mt}.

Despite the progress made, machine learning-based approaches in glass composition prediction and screening face several limitations~\cite{Ren:2018:SA,Hu:2020:npjCM,Mont:2020:IMR,Sarker:2022:APR}. Firstly, existing methods predominantly formulate the problem as a regression task, with a primary focus on predicting T$_g$~\cite{Alcobaça:2020:Acta,Cassar:2018:Acta}. However, accurately fitting the glass transition temperature (T$_g$)  necessitates an extensive dataset for training~\cite{Li:2024:AM}, which is frequently constrained in real-world applications. In addition, samples with T$_g$ values in the 300 - 600°C range display a high frequency, while those from other ranges occur less frequently. Regression tasks, which aim to optimize the squared error across all samples, are often dominated by these high-frequency "head" samples. Consequently, the predictive performance of the model tends to decline significantly in the tail regions. Secondly, most methods rely on supervised learning paradigms that are highly dependent on data quality~\cite{Wu:2023:AM}. Glass databases like Interglad and SciGlass suffer from inconsistencies, numerous outliers, and outdated information. This results in substantial noise within the samples, particularly evident in cases where the proportions of components do not sum up to one ~\cite{Bhattoo:2022:Acta}.Furthermore, the higher-order interactions between individual components that influence glass properties remain largely unexplored. For instance, \ce{Al^{3+}} readily combines with alkali metal ions to form aluminate or tetra-aluminate structures, altering the overall architecture and property of the glass, thereby rendering the relationship between T$_g$ and individual components beyond the scope of first-order correlations ~\cite{Atila:2020:PCCP}. Although current deep learning approaches have made progress in modeling glass properties~\cite{fan:2021:nc}, there remains an opportunity to develop more specialized architectures that can explicitly capture these intricate chemical interactions, potentially leading to improved predictive performance.

\begin{figure}[!t]
	\centering
	\includegraphics[width=0.46\textwidth]{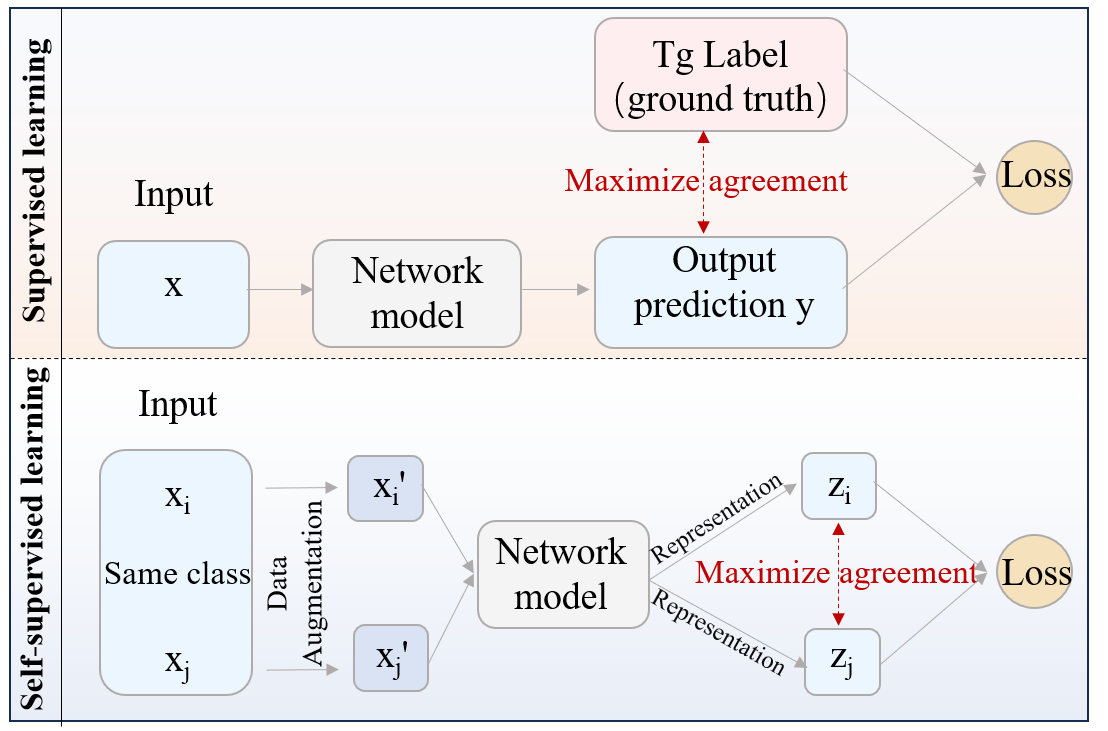}
	\caption{Illustrative example of supervised learning and self-supervised learning. Supervised learning optimizes the consistency between model predictions and labels, while self-supervised learning maximizes representation consistency between augmented views of the same instance through pretext task.}
	\label{Fig:SSL1}
\end{figure}

Self-supervised learning has emerged as a promising direction in artificial intelligence\footnote{\url{https://spectrum.ieee.org/yann-lecun-ai}}, demonstrating highly promising performance in a wide range of tasks, including computer vision~\cite{Sim:ICML:2020}, information retrieval~\cite{Liu:2024:TKDE}, and natural language processing~\cite{Chung:inproceedings}.  It distinguishes itself from supervised learning by designing pretext tasks~\cite{He_2020_CVPR} for model training, as depicted in Fig.~\ref{Fig:SSL1}.
In this study, we design a novel self-supervised learning framework tailored for glass composition screening. This framework holds the potential to be easily generalized to other material discovery tasks. Primarily, we formulate the task of compositional screening as a classification problem, where the model is tasked with directly predicting whether the T$_g$ of a specific composition falls within a predefined range. Subsequently, we propose a data augmentation strategy rooted in asymptotic theory, aiming to both expand the training dataset and enhance its model's robustness against noise. Following this, we design \textit{DeepGlassNet}, a foundational network architecture meticulously crafted to extract and distill the interaction features among the compositional constituents. Ultimately, we integrate a self-supervised learning task, enabling the model to optimize the Area Under Receiver Operating Characteristic Curve (AUC) as a crucial metric for evaluating classification performance.
To facilitate practical applications, we have released both the code and dataset, ensuring compatibility with multiple operating systems and supporting both GPU and CPU computations. Users can seamlessly incorporate their own datasets in the specified format, define the desired properties and their respective ranges, and train customized material screening models tailored for new material designs.

\begin{figure*}[!h]
	\centering
	\includegraphics[width=0.95\textwidth]{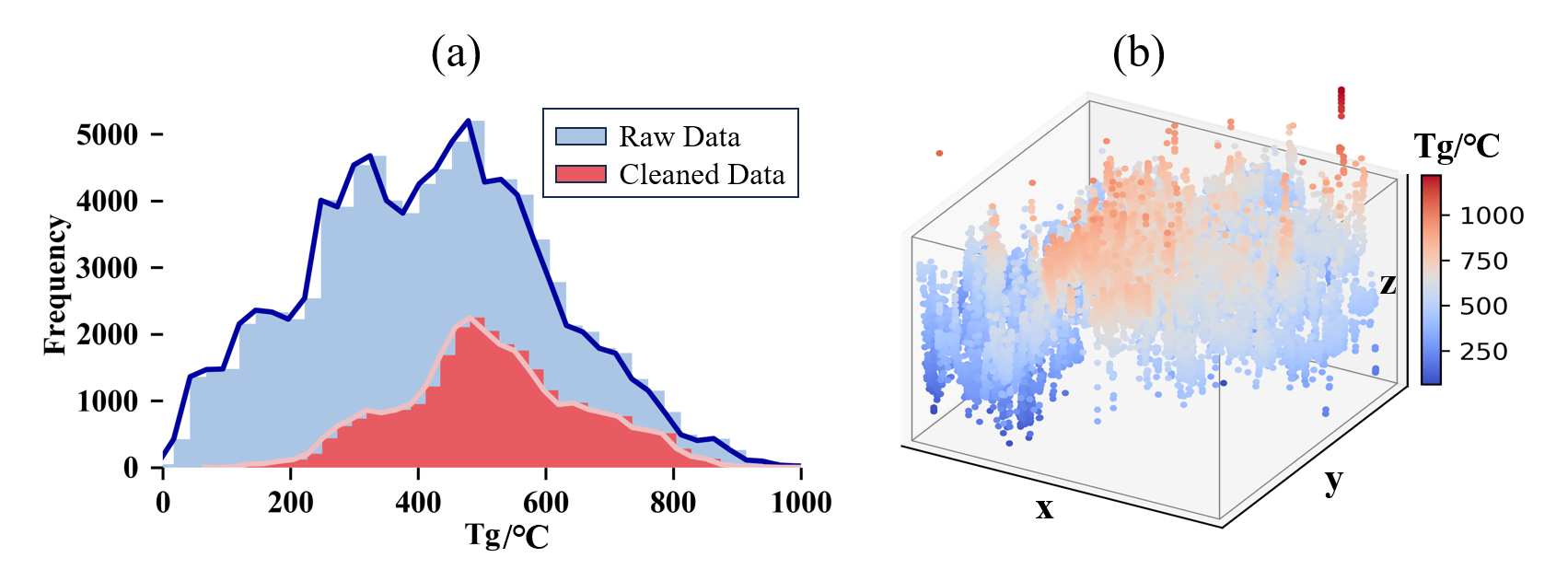}
	\caption{(a) Distributions of raw and cleaned dataset,~(b) TSNE visualization of cleaned dataset: The x- and y-axes display two-dimensional embeddings obtained through TSNE  dimensionality reduction of 18 glass compositional features, while the z-axis corresponds to the target property label T$_g$.}
	\label{Fig:fram}
\end{figure*}

\section{Method}

\subsection{Data Collection and Cleaning}
The dataset utilized in this study was sourced from version 7.12 of the SciGlass database, which compiles comprehensive data from scientific journals, books, and patents~\footnote{\url{https://github.com/epam/SciGlass}}. The initial dataset comprised approximately 422,000 glasses and melts, including more than 268,000 oxide glasses and melts. However, substantial measurement or clerical errors resulted in numerous glass compositions where the mass fractions failed to sum to unity.  Within an acceptable error range, we filtered the samples to include only those with total mass fractions between [0.95, 1.05] and corresponding T$_g$ labels, excluding others to mitigate the risk of the model learning incorrect patterns from excessive noise. 
We initially removed inconsistent and incomplete entries, followed by the elimination of duplicate samples. The compositional data were rounded to five decimal places and restricted to 18 major oxides: \ce{SiO2},\ce{Al2O3},\ce{B2O3},\ce{CaO},\ce{Na2O},\ce{Li2O},\ce{MgO},\ce{BaO},\ce{ZnO},\ce{K2O},\ce{PbO},\\ \ce{SrO},\ce{GeO2},\ce{ZrO2},\ce{TiO2},\ce{TeO2},$\text{Fe}_\text{m}\text{O}_\text{n}$ and \ce{P2O5}. This means each sample's composition consists of one or more of these 18 components. The 18 components were selected because they comprehensively represent essential structural roles in oxide glasses, while covering most of industrially and scientifically relevant compositions. This ensures sufficient data density for robust ML modeling.This procedure yielded a cleaned dataset containing about 35176 samples. Fig.~\ref{Fig:fram}~(a) exhibits the histograms of T$_g$ values in both the original and cleaned datasets. It is evident that the majority of samples concentrate within the temperature range of 300-600°C. When the histograms are arranged in descending order, a typical long-tail distribution emerges. Regression methods, exemplified by mean squared error loss, tend to overfit the high-frequency "head" samples while underfitting the low-frequency "tail" samples. Fig.~\ref{Fig:fram}~(b) presents the t-SNE visualization of the cleaned dataset, where the x-axis and y-axis correspond to the coordinates obtained after t-SNE dimensionality reduction of the 18 components fractions, and the z-axis corresponds to the T$_g$ labels. The visualization reveals that the data does not exhibit clear clustering or linearly separable features, indicating the complexity of the relationship between compositions and T$_g$.

Subsequently, the cleaned dataset was randomly divided into a training set (80\%) and a validation set (20\%), consisting of 28153 samples for training and 7023 samples for evaluation, respectively. Our test composition selection enforced fundamental physicochemical constraints—requiring network formers >50 wt\% and modifiers <50 wt\% to ensure glass-forming ability—while implementing combinatorial optimization through nine dominant oxides at 5 wt\% resolution increments. This approach generated about 39k stoichiometrically balanced compositions ($\sum$=100 wt\%) that maintain physicochemical realizability and computational tractability. The synthetic  test set was used for material discovery instead of  model evaluation. It enabled practical composition screening—identifying novel, experimentally untested glass candidates within target T$_g$ ranges. This aligns with our core objective: accelerating discovery beyond existing datasets.  
This approach is well-established in materials informatics for inverse design, where synthetic spaces uncover promising candidates absent from known databases~\cite{schwalbe:2020:ML}.

\subsection{Label Transformation}
Composition screening is fundamentally a classification task, aimed at classifying samples with desirable T$_g$ values from a potential set of compositions to minimize trial-and-error in the discovery of new materials. Based on this reason, we predefine a target T$_g$ range, such as 500-600°C (adjustable as necessary). Samples that fall within this range are designated as target samples(target class), while those outside are considered non-target samples(non-target class).For target samples, their T$_g$ labels are assigned a value of 1, whereas non-target samples are labeled as 0. This transformation shifts the focus from predicting the exact T$_g$ value of samples in the test set to predicting whether the T$_g$ of samples falls within the specified range. The advantages of this approach are two-fold: Firstly, it eliminates T$_g$ noise, thereby preventing the model from overfitting to incorrect T$_g$ labels. Secondly, compared to a regression task that accurately fits T$_g$ values, classification is inherently simpler~\cite{oord2018representation}, reducing the demands on model fitting capability and the amount of training data required for the model.

\begin{figure*}[!h]
	\centering
	\includegraphics[width=\textwidth]{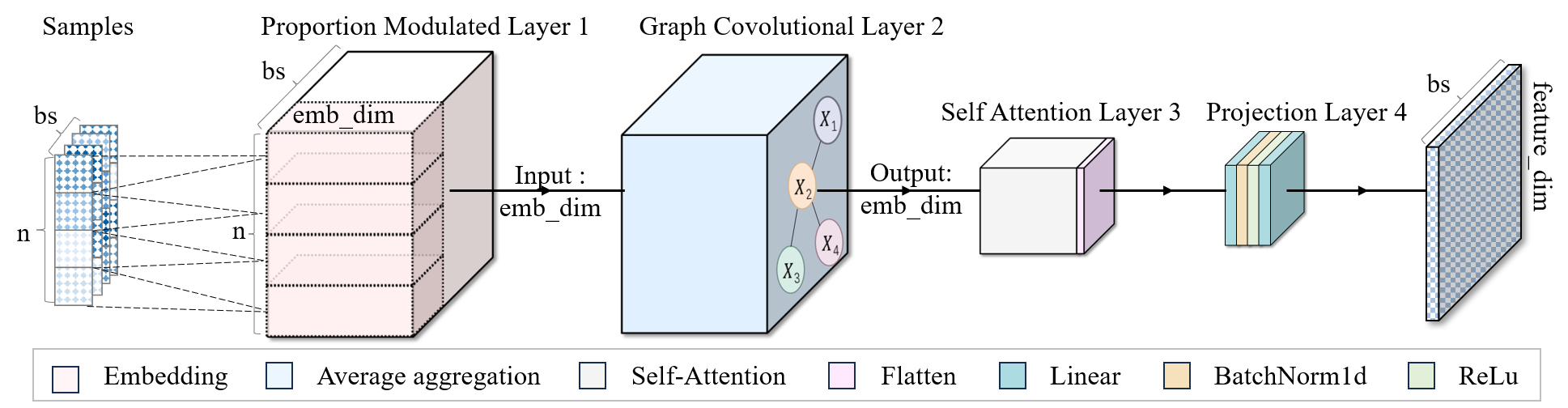}
	\caption{Machine learning model architecture. The dimensions of each module are denoted by the following symbolic conventions: bs represents batch size, n denotes the number of compositional elements, emb\_dim signifies embedding dimension, and feature\_dim indicates output feature dimension.}
	\label{Fig:md}
\end{figure*}
\subsection{Data Augmentation}
The fundamental principle of data augmentation involves introducing slight perturbations to the samples while maintaining their semantic information. It originated in the field of computer vision. Taking an image of a dog as an illustration, after undergoing typical data augmentation techniques, such as random cropping, rotation, and Gaussian blurring, humans can still readily identify it as a dog. This approach not only effectively expands the dataset but also enhances the model's robustness against noise. In this section, we will delve into the specific data augmentation strategies tailored for materials discovery tasks.

Existing image augmentation techniques are not directly applicable to glass composition data. Due to the diverse experimental environments and conditions from which experimental datasets often originate, errors in measurements are inevitable. We argue that, for a given componential ratio, two independent measurements resulting from measurement errors should possess identical properties, implying that the semantics of the original sample and the augmented sample subjected to small error perturbations remain unchanged.

It is essential to observe that the total error in measuring componential ratios is the sum of various errors, typically including sampling errors stemming from microscale inhomogeneity of the samples, instrumental errors, observational errors, and so on. Assuming these errors are independently and identically distributed, according to the central limit theorem, the total error asymptotically approaches a normal distribution. On this basis, we apply small perturbations with a normal distribution to each component of the glass composition to emulate error perturbations for data augmentation purposes:
\begin{eqnarray}\label{eq:dataaug}
	x_i' = x_i \times (1 + \epsilon), ~~ \epsilon \sim \mathcal{N}(0,\sigma^2) 
\end{eqnarray}
where \( x_i \) represents the original proportion of component \( i \in [1,2,\cdots,n]\), and $n$ is the number of components. \( \epsilon_i \) is a random value drawn from a normal distribution \( \mathcal{N}(0, \sigma^2) \), where \( \sigma \) is a small standard deviation representing measurement noise. Note that $\sigma$ functions not as a fixed physical constant,  but rather as a tunable hyperparameter that exploits the asymptotic error structure. By adjusting the data perturbation intensity during augmentation through $\sigma$, we aim to encourage the model to learn sample features independent of noise, thereby enhancing the model's robustness. Mathematically, when $x_i$=0, $x_i^\prime=0×(1+ \epsilon)=0$ regardless of $\epsilon$. Thus, while perturbations are technically applied universally, zero components remain unchanged. This design aligns with physical realism (zero components stay absent) while maintaining augmentation consistency across all dimensions.

In line with the principles of self-supervised learning, perturbations are applied only during the training phase. This encourages the model to produce consistent representations from augmented views of each sample, thereby strengthening its ability to capture high-level semantic features rather than low-level noise. During inference, however, neither the validation nor the test sets undergo any perturbation. This ensures that the evaluation is conducted on original, unmodified data that reflect realistic measurement conditions, thereby providing a reliable assessment of the model's generalization performance.

\subsection{Data Normalization}
The concentrations of different glass components vary widely, leading to significant differences in their magnitudes. To address this, we normalized the composition data using Z-Score normalization for each component proportion, \( x_i \), as follows:
\begin{eqnarray}\label{eq:datanorm}
	x_i^\prime = \frac{x_i - \bar x_{i}}{\text{std}(x_i)}
\end{eqnarray}
where $ \bar{x}_{i} $ and $ \text{std}(x_i) $ represent the mean and standard deviation of the compositional ratio $ x_i $. All the samples in the training set, validating set, and test set were normalized before being fed into the model. 
	Following Z-score normalization, the inherent sparsity of the original composition data (i.e., exact zero values indicating the absence of a component) was not preserved, as absent components are transformed to negative values.

\subsection{Machine Learning Model Architecture}

Glass is composed of multiple components, each exerting a distinct influence on the glass transition temperature (T$_g$). The impact of a single component on material properties is termed the ‌\textbf{first-order} effect‌. For example,  \ce{SiO2} enhances T$_g$ by forming \ce{SiO4} tetrahedron~\cite{Cassar:2018:Acta}. Furthermore, this component interacts with others to form special structures, which in turn produce a ‌\textbf{second-order‌} or ‌high-order effect‌ on the material properties. In the case of aluminosilicate glass, when \ce{Al^{3+}} enters the glass network as \ce{AlO4} tetrahedron, it reinforces the \ce{SiO4} tetrahedron connected by shared oxygen atoms. This reinforcement strengthens the network structure, leading to an increase in T$_g$. However, \ce{Al^{3+}} may also combine with alkali metal ions to form aluminate structures. These structures can alter the overall architecture and characteristics of the glass, rendering the impact on T$_g$ uncertain and non-linear~\cite{Atila:2020:PCCP}. In this section, we introduce \textit{DeepGlassNet}, an artificial neural network capable of capturing the complex first-order and high-order  effects of components on properties, in order to extract optimal feature representations of glass properties. The model comprises four key modules: proportion-modulated embedding, interaction feature extraction, self-attention, and nonlinear projection, as shown in Fig.~\ref{Fig:md}.

\subsubsection{Proportion Modulated Embedding Layer}
To capture the \textbf{first-order effect} of components on properties, we introduce a proportion modulated embedding layer, which adjusts the embedding of each component based on its proportion within the glass composition.
Given a glass composition with $n$ components, each with a proportion \( x_i \) (where \( i = 1, 2, ..., n \), \(\sum_{i=1}^{n} x_i = 1\)), we associate each component with a learnable embedding vector \( \mathbf{e}_i \in \mathbb{R}^d \), where \( d \) is the embedding dimension. The embedding is modulated by the component proportion thro-ugh a linear transformation:
\begin{eqnarray}\label{eq:propomodu}
	\mathbf{h}_i = x_i \mathbf{e}_i
\end{eqnarray}
Here, \( \mathbf{h}_i \) represents the modulated feature for the \( i \)-th component. This modulation allows the model to directly incorporate proportion information into the embedding, ensuring that the representation reflects the varying impact of different glass component proportions. 
\begin{figure*}[!h]
	\centering
	\includegraphics[width=0.95\textwidth,height=0.3\textwidth]{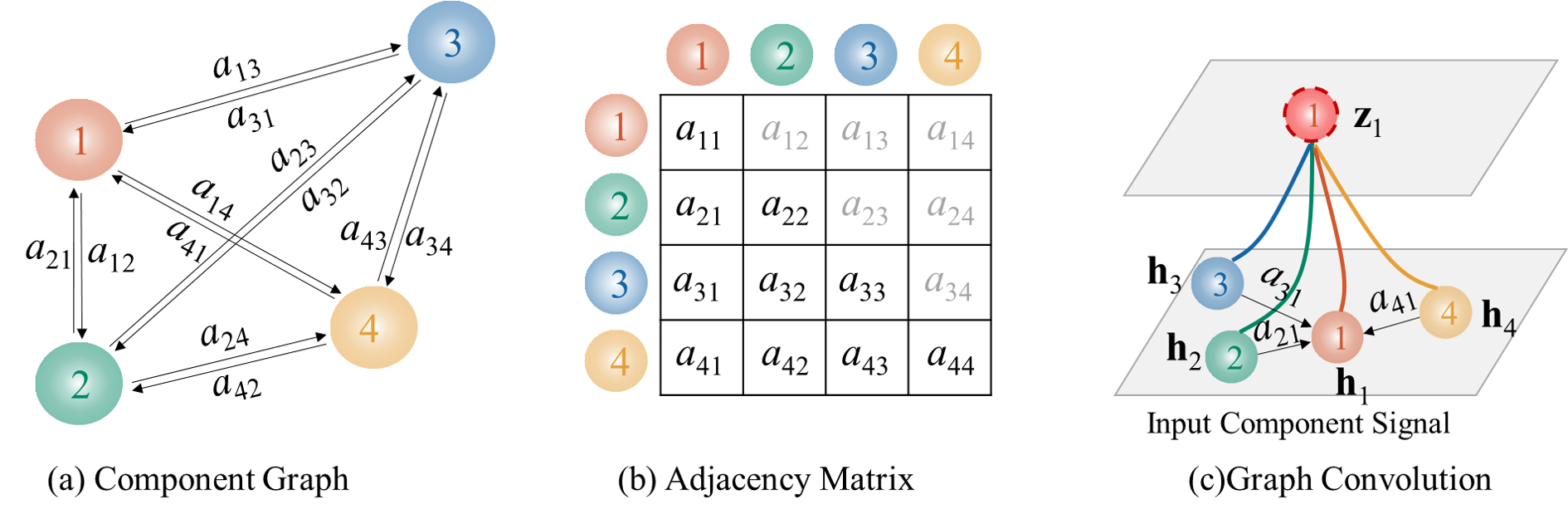}
	\caption{Illustrative Explanation of Graph Convolution(Demonstrated on four glass components labeled 1–4). (a) Component Graph. Nodes represent glass components. Edge weight $a_{ij}$ quantifies the interaction strength between components $i$ and $j$, $i,j \in \{1,2,3,4\}$. (b) Adjacency Matrix. Matrix representation of the component graph. Each element $a_{ij}$ in $\mathbf{A}$ explicitly encodes the
		pairwise interaction strength between component $i$ and component $j$. (c) Graph Convolution Mechanism for node 1: Neighboring nodes (2, 3, 4) propagate their input features $\mathbf{h}_2, \mathbf{h}_3, \mathbf{h}_4$ to node 1, weighted by interaction strengths $a_{21}, a_{31}$ and $a_{41}$. The output feature $\mathbf{z}_1$ is generated through weighted aggregation of these signals with node 1’s intrinsic feature $\mathbf{h}_1$.}
	\label{Fig:so}
\end{figure*}

\subsubsection{Graph Convolution Layer}
To capture the \textbf{second-order effect} of components on
properties, we introduce a graph convolution layer. The core operation represents each component as a node in a graph, with edges signifying interactions between components. 
Fig.~\ref{Fig:so}(a) illustrates this concept using four glass components (labeled 1–4), where nodes represent glass  components, edge weight $a_{ij}$ quantifies the interaction strength between components $i$ and $j$, 
where $i,j \in \{1,2,3,4\}$.In the context of $n$ components, the component graph  is formalized by its adjacency matrix $\mathbf{A}\in \mathbb{R}^{nxn}$(Fig.~\ref{Fig:so}(b)). 
Each element $a_{ij}$ in $\mathbf{A}$ explicitly encodes the
pairwise interaction strength between component $i$ and component $j$. 

Then the adjacency matrix is parameterized as the product of two low-rank matrices \( \mathbf{V} \in \mathbb{R}^{n \times f} \), where \( n \) is the number of components and \( f \) is the dimension of the latent space used to capture the interactions:
\begin{eqnarray} \label{eq:adj}
	\mathbf{A} = \mathbf{V} \mathbf{V}^\mathsf T \in \mathbb{R}^{n \times n} 
\end{eqnarray}
Note, \(\mathbf{V}= [\mathbf{v}_1,\mathbf{v}_2, \cdots, \mathbf{v}_n]^\mathsf T \), where $\mathbf{v}_i \in \mathbb R^f$ represents the normalized latent vector for component \( i \). This low-rank factorization of the adjacency matrix ensures that the adjacency matrix is a real symmetric matrix~\cite{FM:ICDM:2010}
\begin{eqnarray}
	\mathbf{A}^\mathsf T = (\mathbf{V} \mathbf{V}^\mathsf T)^\mathsf T=\mathbf{V} \mathbf{V}^\mathsf T = \mathbf{A}
\end{eqnarray}
The condition $\mathbf{A}^\mathsf{T} = \mathbf{A}$ implies that each element satisfies \( a_{ij} = a_{ji} \). This equality indicates that the interaction strength between components $i$ and $j$ is symmetric—that is, the effect of $i$ on $j$ is identical to that of $j$ on $i$. Such symmetry is consistent with fundamental physical principles, which often require reciprocity and equivalence in interactions among components within a system. Moreover, the use of low-rank factorization is closely linked to parameter efficiency and inherent symmetry. It not only ensures that the adjacency matrix remains symmetric but also reduces the number of parameters to be learned from \( \binom{n}{2} = n(n-1)/2 \) to \( n \times f \), where $f$ is a small fixed dimension. This reduction in parameters becomes particularly significant when $n$ is large.

The graph convolution operation modulates each component’s embedding \( \mathbf{h}_i \), which was initially derived from the proportion-modulated embedding layer, using the interaction information from the adjacency matrix:
\begin{eqnarray}\label{eq:graph}
	\mathbf{z}_i = \mathbf{h}_i+ \frac{1}{n-1} \sum_{j \neq i} a_{ij} \mathbf{h}_j , ~~i \in \{1,2,\cdots, n\}. 
\end{eqnarray}
This operation can be viewed as a message-passing mechanism where each component \( i \) receives information from its neighboring components $j$, weighted by the second-order interaction strength of component $a_{ij}$. The result, \( \mathbf{z}_i \), is a refined embedding for component \( i \) that incorporates second-order interaction information from other components in the composition.  
Noticing that \( \mathbf{V}_i \) 
is a normalized vector, it follows that \( \mathbf{V}_i \mathbf{V}_i^T=1  \)
ensuring that the main diagonal elements of $\mathbf{A}$ are equal to 1. So, each component’s representation ($\mathbf{z}_i$) combines its own features ($\mathbf{h}_i$) with a weighted average of features from interacting partners ($\mathbf{h}_j$), where weights $a_{ij}$ are learned from data. Fig.~\ref{Fig:so}(c) provide a illustrative example of graph convolution mechanism for node 1: Neighboring nodes (2, 3, 4) propagate their input features $\mathbf{h}_2, \mathbf{h}_3, \mathbf{h}_4$ to node 1, weighted by interaction strengths $a_{21}, a_{31}$ and $a_{41}$. The output feature $\mathbf{z}_1$ is generated through weighted aggregation of these signals with node 1’s intrinsic feature $\mathbf{h}_1$.

The graph convolution operation offers several key advantages: 1. Second-Order Interaction Modeling. By modulating each component’s embedding using the graph structure, the model explicitly captures second-order interactions between components. This allows the model to learn how different combinations of components influence the overall glass properties.
2. Efficient Parameterization: The adjacency matrix 
$\mathbf{A}$ is parameterized by the product of low-rank matrices 
$\mathbf{V}$ (e.g., with rank $f=4$). Parameterizing $\mathbf{A}$ in this way requires only $n×f$ parameters to be learned. This approach not only guarantees symmetric interaction strengths between components $i$ and $j$, but also drastically reduces the number of parameters to be learned compared to directly parameterizing the adjacency matrix  $\mathbf{A}$, which would require $n×n$ parameters. This operation allows the model to handle intricate glass compositions with numerous interacting elements.
3. Rich Feature Representation: The embeddings produced by the graph convolution layer are enriched by the structural information encoded in the adjacency matrix. The model is able to distill important interaction information and produce feature representations that are not only informed by each component’s individual properties but also by the critical interactions that shape the material's overall behavior. This makes the model well-suited for predicting the complex properties of glass compositions.
\subsubsection{Self-Attention Layer} 
To enable the model to focus on specific components or structures that have a significant impact on material properties, we introduce the self-attention mechanism. Furthermore, it excels at capturing complex, high-order interactions between components, rendering it an ideal approach for refining the feature representations obtained from earlier layers. Given the refined embeddings \( \mathbf{z}_i \) in Eq.~\eqref{eq:graph} from the graph convolution layer, the self-attention mechanism computes three matrices: query \( \mathbf{Q} \), key \( \mathbf{K} \), and value \( \mathbf{V} \), where:
\begin{eqnarray} \label{eq:keyquery}
	\mathbf{Q} = \mathbf{W}_Q \mathbf{z}_i, \quad \mathbf{K} = \mathbf{W}_K \mathbf{z}_i, \quad \mathbf{V} = \mathbf{W}_V \mathbf{z}_i
\end{eqnarray}
Here, \( \mathbf{W}_Q \), \( \mathbf{W}_K \), and \( \mathbf{W}_V \) are learnable weight matrices that transform the input embeddings into query, key, and value vectors. The attention score between components \( i \) and \( j \) is computed using the scaled dot product of the query and key vectors:
\begin{eqnarray}\label{eq:attentionweight}
	\alpha_{ij} = \text{softmax}(\frac{\mathbf{Q}_i \cdot \mathbf{K}_j^\mathsf T}{\sqrt{d_k}})
\end{eqnarray}
where \( d_k \) is the dimensionality of the key vectors. The attention scores \( \alpha_{ij} \) are then used to compute a weighted sum of the value vectors \( \mathbf{V}_j \), generating a new embedding for each component:
\begin{eqnarray} \label{eq:attvalue}
	\mathbf{u}_i = \sum_{j=1}^{n} \alpha_{ij} \mathbf{V}_j
\end{eqnarray}
This operation can be seen as learning a high-order interaction map between components, where each component’s final representation \( \mathbf{u}_i \) is influenced by all other components in the glass composition, weight-ed by their learned relevance. 

\subsubsection{Nonlinear Projection Layer}

The self-attention layer outputs a feature tensor \( \mathbf{U} \in \mathbb{R}^{n \times d} \) of n components, which is flattened into a vector \( \mathbf{u}_\text{flat} \in \mathbb{R}^{n \cdot d} \). we apply a projection head $g(\cdot)$ to map this vector into a lower-dimensional space \( \mathbb{R}^k \).
Projection heads have demonstrated remarkable performance in artificial intelligence tasks, especially in the field of computer vision~\cite{Sim:ICML:2020}, where they play a crucial role in creating compact and meaningful feature representations.

Specifically, the projection head comprises a four-layer fully connected neural network. The first layer reduces the dimensionality of the input vector, followed by batch normalization and a ReLU activation function. The second layer, which is linear, maps the features to the desired output dimension. An additional normalization step is incorporated to ensure the stability of the predictions:
\begin{eqnarray}
	\mathbf{f} &=& g(\mathbf{u}_\text{flat} ) \in \mathbb{R}^k\nonumber \\
	&=&  \mathbf{W}_2 \cdot \text{ReLU}(\mathbf{W}_1 \mathbf{u}_\text{flat} + \mathbf{b}_1) + \mathbf{b}_2
\end{eqnarray}
Here, \( \mathbf{W}_1 \in \mathbb{R}^{n \cdot d \times h} \) and \( \mathbf{W}_2 \in \mathbb{R}^{h \times k} \) are learnable weight matrices, where \( h \) is the hidden dimension, and \( k \) is the output feature dimension.

\subsection{Self-supervised Learning}
\begin{figure*}[!h]
	\centering
	\includegraphics[width=\textwidth,height=0.45\textwidth]{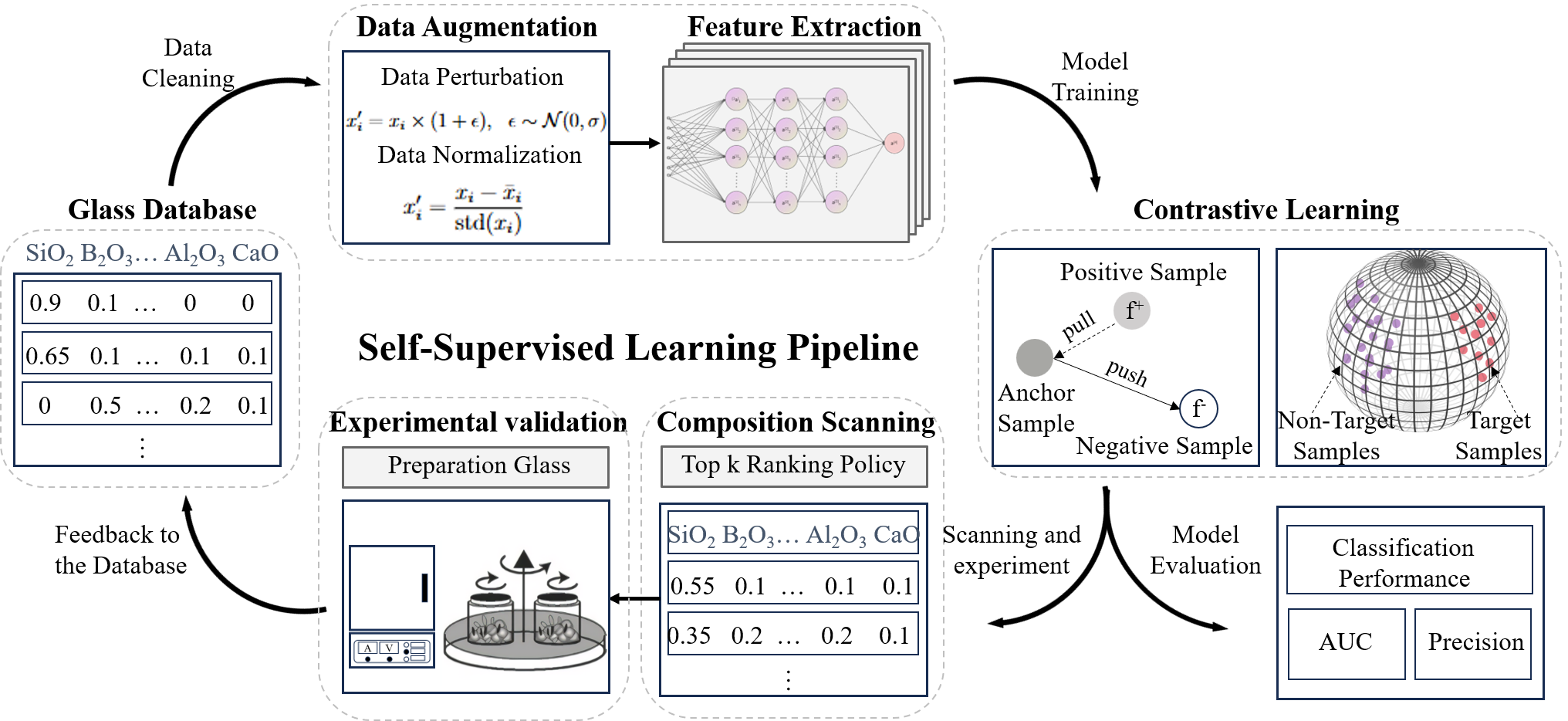}
	\caption{Self-supervised learning pipline.}
	\label{Fig:flow}
\end{figure*}
Self-supervised learning employs self-supervised signals to supervise the model's output, thereby relieving the model from accurate reconstructing of T$_g$ labels. For samples belonging to the same class, given their identical class semantics, our objective is to maximize the consistency of their feature representations. Conversely, for samples from different classes, due to their distinct class semantics, we strive to minimize the consistency of their feature representations. Specifically, for any given sample $\mathbf{x}$, we select a sample from its same-class counterparts as a positive sample $\mathbf{x}^+$ and another from different-class samples as a negative sample $\mathbf{x}^-$, organizing the data into a triplet form $(\mathbf{x}, \mathbf{x}^+, \mathbf{x}^-)$. Subsequent to data augmentation and feature extraction, we procure the feature representations of the triplet $(\mathbf{f}, \mathbf{f}^+, \mathbf{f}^-)$ and proceed to compute the following contrastive loss:
\begin{eqnarray}\label{eq:loss}
	\mathcal{L} = - \mathbb{E}_{(x,x^+,x^-)}\log \frac{\exp
		\langle \mathbf{f}, \mathbf{f}^+ \rangle}{\exp
		\langle\mathbf{f}, \mathbf{f}^+\rangle + \exp
		\langle\mathbf{f}, \mathbf{f}^-\rangle}
\end{eqnarray}
where $	\langle \mathbf{f}, \mathbf{f}^+ \rangle$ is the inner product similarity for two vectors $ \mathbf{f}=[{f}_1,{f}_2,...,{f}_k]$ and $  \mathbf{f}=[{f}^+_1, {f}^+_2,..., {f}^+_k]$, it is defined as: $\langle \mathbf{f}, \mathbf{f}^+ \rangle= \sum_{i=1}^k f_i\cdot f^+_i$. So, $\langle\mathbf{f},\mathbf{f}^+\rangle$ measures the similarity between   anchor feature $\mathbf{f}$ and positive sample feature $\mathbf{f}^+$,  $\langle\mathbf{f},\mathbf{f}^-\rangle$ measures the similarity between  anchor feature $\mathbf{f}$ and negative sample feature $\mathbf{f}^-$. The minimum of the loss is 0, which is attained if and only if $\exp\langle \mathbf{f}, \mathbf{f}^+ \rangle\rightarrow +\infty$ and $\exp
\langle \mathbf{f}, \mathbf{f}^- \rangle\rightarrow -\infty$. In this case, maximum consistency is achieved among samples of the same class, whereas minimum consistency is achieved among samples from different classes.

This learning paradigm exhibits several advantages. Firstly, for each anchor, the inclusion of both a positive and a negative sample in the computation assists in mitigating class imbalance issues and prevents the model from overfitting to the majority class. Additionally, this learning paradigm relieves the model from reconstructing T$_g$ labels~\cite{oord2018representation}. Thirdly, directly optimizing this loss leads to the optimization of the classification metric AUC~\cite{Liu:2024:TKDE}. 

\subsection{Model Training, Evaluation, and Screening}
The dataset, data preprocessing, model training, validation, and screening codes are publicly accessible, allowing users to execute the entire workflow simply by running the \verb|main.py| script.  This code is implented in PyTorch, a state-of-the-art deep learning framework, to execute all computations using tensor operations. Training, validation, and testing datasets were structured as \verb|Dataset| class to enable efficient GPU batch processing. The overall workflow is depicted in Fig.~\ref{Fig:flow}, which consists of three key steps:
\begin{itemize}
	\item Training: Train \textit{DeepGlassNet} using the samples from the training set. The network learns to extract meaningful feature representations by optimizing the loss through self-supervised learning task.
	
	\item Evaluation: Evaluate the model's performance using the validation set and fine-tune hyperparameters to optimize classification accuracy. This step ensures that the model generalizes well to unseen data.
	
	\item Screening: Use the trained model to scan other potential compositions (test set) and select the most promising samples that are likely to meet the desired criteria for material preparation.
\end{itemize}
The following sections describe the principles underlying the training, validation, and screening phases. 

\subsubsection{Training}
For each sample \( \mathbf x \) in the training set, a positive example and a negative example were randomly selected based on the label, which indicates whether the sample is a target sample or a non-target sample. These samples were organized as training triplet \((\mathbf x, \mathbf x^+, \mathbf x^-)\), subsequently undergoing data augmentation and normalization before being fed into \textit{DeepGlassNet}. The network output was a \( k \)-dimensional feature triplet \((\mathbf f, \mathbf f^+, \mathbf f^-)\), and the loss was calculated based on Eq.~\eqref{eq:loss}. The model parameters of DeepGlassNet were optimized using the Adam optimizer.

\begin{figure}[!h]
	\centering
	\includegraphics[width=0.4\textwidth]{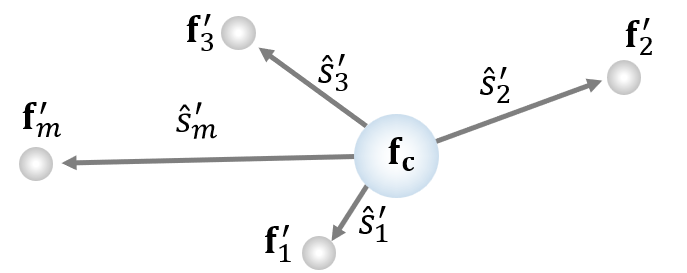}
	\caption{Illustrative example of model evaluation. $\mathbf{f}_\text{c} $ denote the feature space centroid of target-class instances in the training set,
		$\mathbf{f}^\prime_m$ represents the feature vector of the $m$-th sample in the validation set,
		$\hat{s}^\prime_m$ computes their inner product similarity,  which subsequently determines the likelihood of the sample belonging to the target class.}
	\label{Fig:f}
\end{figure}
\subsubsection{Valiatation/Evaluation}
Evaluation is conducted utilizing validation data to assess the performance of \textit{DeepGlassNet}, followed by fine-tuning of hyperparameters to attain optimal performance. This process can be regarded as a generalized k-nearest neighbors (k-NN) approach. Specifically, the training process terminates once the training loss converges to a stable and low value. At this juncture, all target samples from the training set are fed into \textit{DeepGlassNet} to obtain their respective feature representations. The mean of these representations, denoted as 
$\mathbf{f}_\text{c}$, was computed and used as the class center for the target class.

Subsequently, samples in validation set are normalized (without augmentation) and fed into the trained \textit{DeepGlassNet} model to obtain their feature representations.
For each of the \( m \) samples in validation set, $m$ similarity scores of validation sample features $\mathbf{f}^\prime_1, \mathbf{f}^\prime_2, \cdots, \mathbf{f}^\prime_m$ with the target class center $\mathbf{f}_\text{c}$ were computed using follow inner product similarity:
\begin{eqnarray}
	[\hat{s}^\prime_1,\hat{s}^\prime_2, \cdots, \hat{s}^\prime_m] = [ \langle\mathbf{f}_\text{c}, \mathbf{f}^\prime_1\rangle, \langle \mathbf{f}_\text{c},\mathbf{f}^\prime_2\rangle, \cdots, \langle\mathbf{f}_\text{c},\mathbf{f}^\prime_m\rangle ]
\end{eqnarray}

As illustrated in Fig.~\ref{Fig:f}, the similarity between the representation of each sample in validation set and the target class center $\mathbf{f}_\text{c}$ reflects the likelihood of that sample belonging to the target class. Ideally, samples in validation set that belong to the target class should exhibit higher similarity to $\mathbf{f}_\text{c}$ than those belonging to the non-target class. To assess the classification performance, the AUC (Area Under Curve) metric was employed: 
\begin{eqnarray}
	\text{AUC} = \frac{\sum_{i=1}^{m_1} \sum_{j=1}^{m_0} \mathbb I(s^\prime_i > s\prime_j)}{m_1 \times m_0},
\end{eqnarray}
where \( \mathbb I(s_i > s_j) \) is an indicator function that takes the value 1 if the similarity score \( s_i \) of a target sample is greater than that of an non-target sample \( s_j \), otherwise 0. \( m_1 \) represents the number of target samples, and \( m_2 \) represents the number of non-target samples. An AUC of 1 implies perfect classification, whereas an AUC of 0.5 corresponds to random guessing.
\begin{figure*}[!h]
	\centering
	\includegraphics[width=0.49\textwidth]{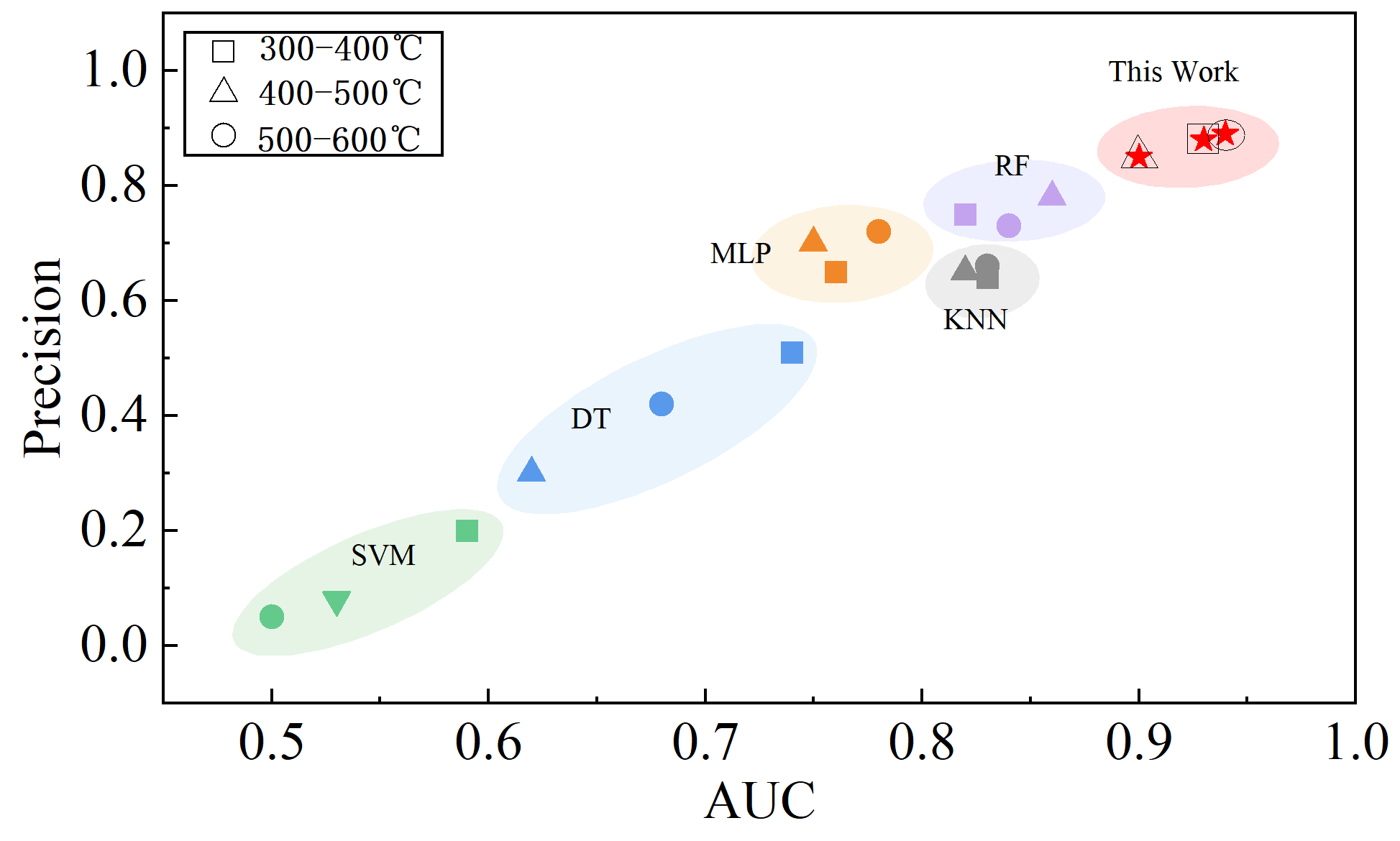}
	\includegraphics[width=0.49\textwidth]{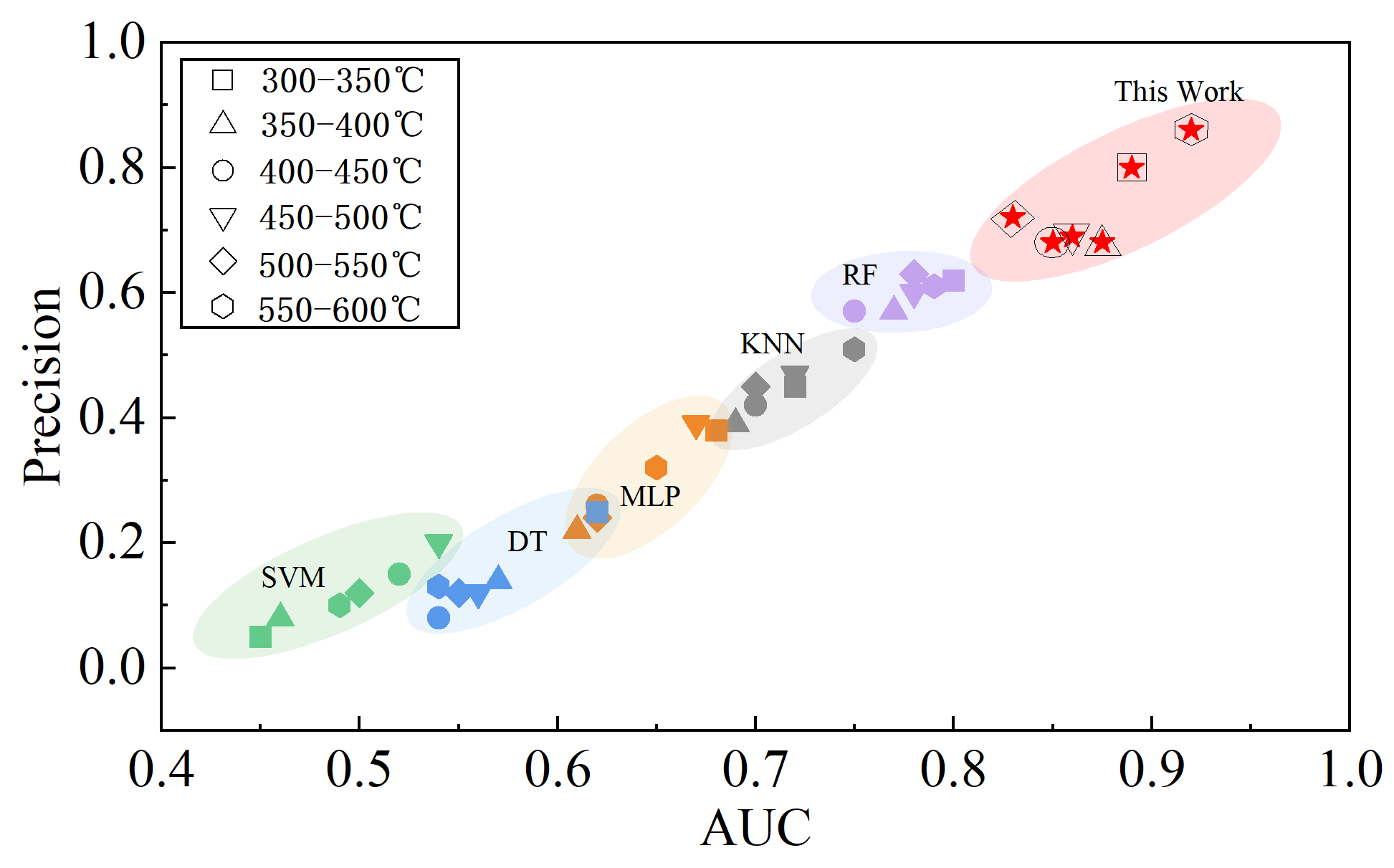}
	\includegraphics[width=0.49\textwidth]{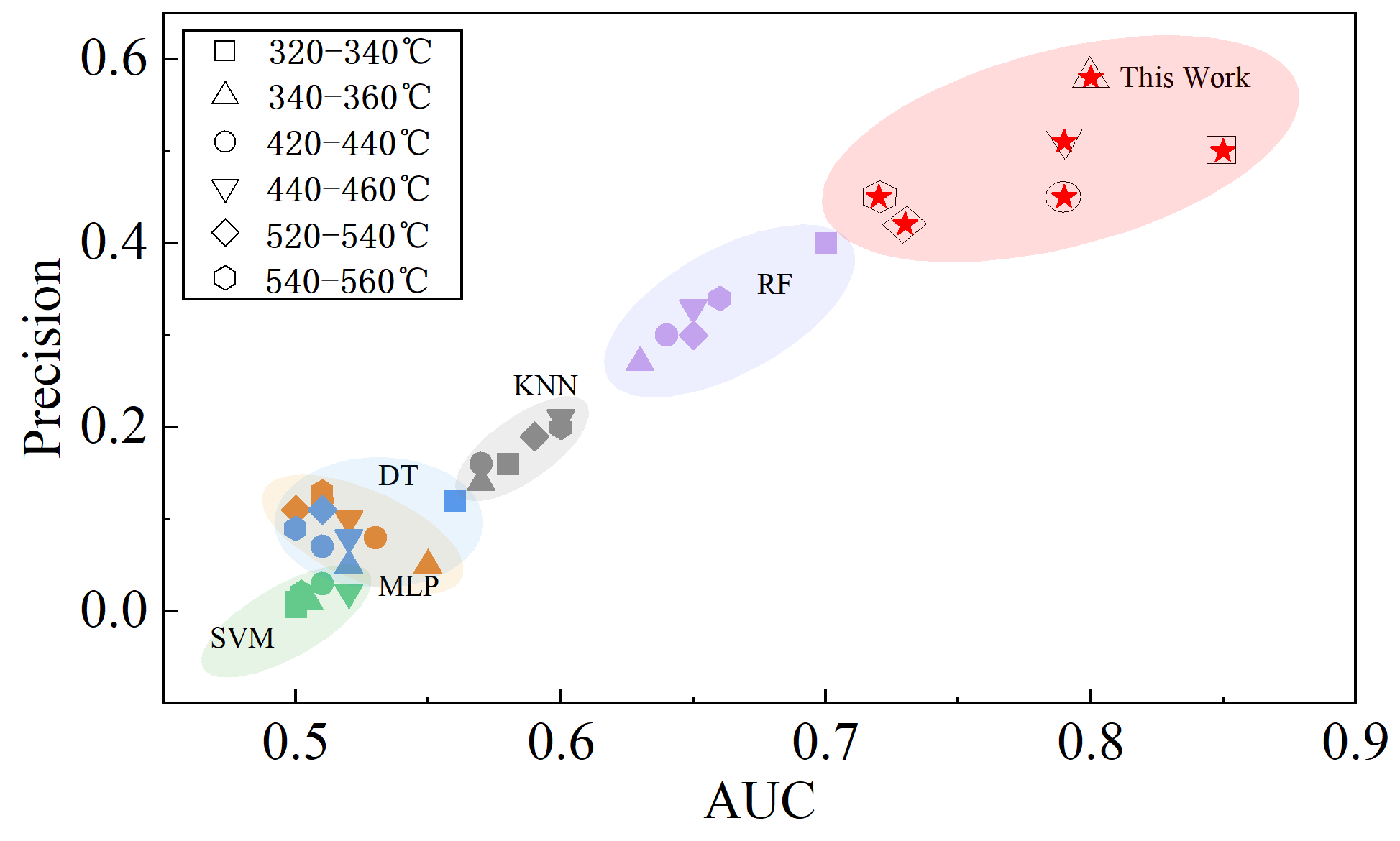}
	\includegraphics[width=0.49\textwidth]{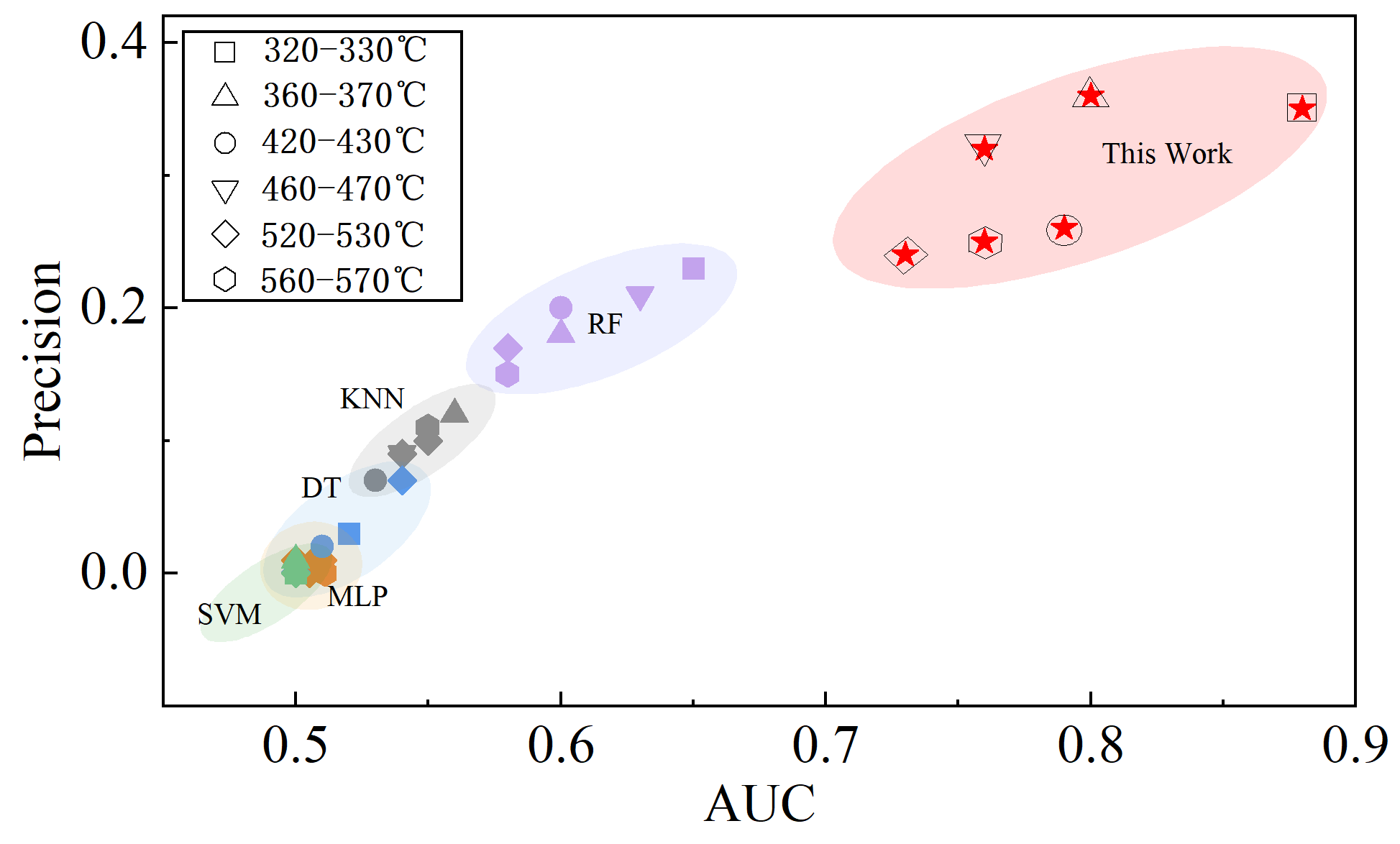}
	\caption{Performance comparison of validation dataset between DeepGlassNet model and KNN,MLP,DT,RF and SVM models in different temperature intervals: The x-axis denotes AUC (Area Under ROC Curve), while the y-axis represents Precision@100.}\label{Fig:TW1}
\end{figure*}

Another critical metric for material screening is precision, particularly for the top-ranked \( k \) samples:
\begin{eqnarray}
	\text{Precision@k} = \frac{\sum_{i=1}^{k} y_i}{k} 
\end{eqnarray}
\begin{figure*}[!h]
	\centering
	\includegraphics[width=0.98\textwidth,height = 0.6\textwidth]{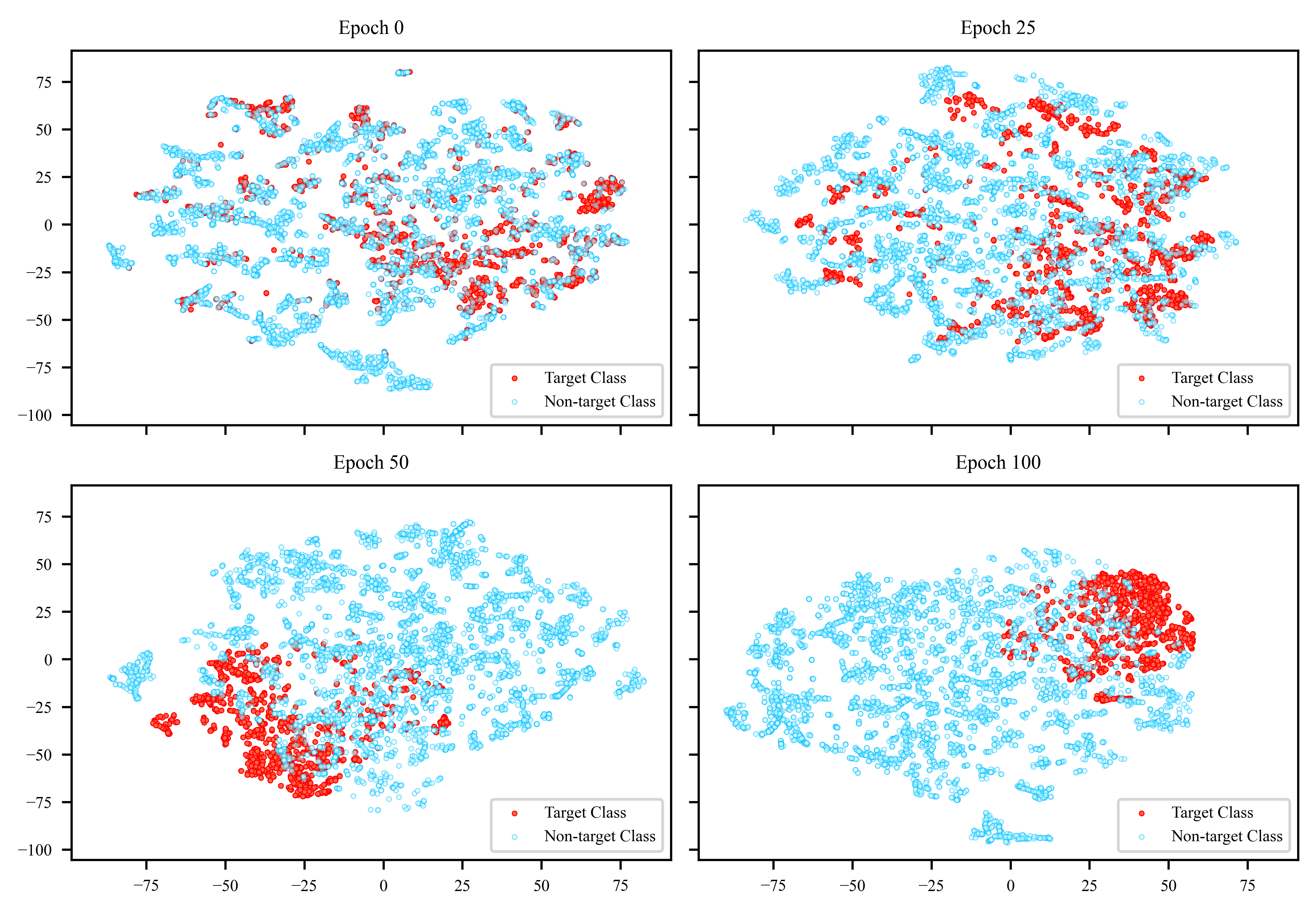}
	\caption{TSNE visualization of the feature representations learned by \textit{DeepGlassNet} on validation set at different training stages within the target temperature range of 500–600°C.}
	\label{Fig:visual}
\end{figure*}
In material composition screening, our objective is to identify the top $k$ samples that have the highest likelihood of belonging to the target class, where ($k$) is a small constant, such as 10, representing the number of samples that can be feasibly prepared. We ranked the samples in the validation set based on their similarity to the target class center and selected the top $k$ samples accordingly. Precision@$k$ is employed to evaluate the ranking accuracy of \textit{DeepGlassNet} among these top-ranked samples. Precision@k assesses \textit{DeepGlassNet}'s classification accuracy among the top-ranked samples. A precision value of 1 indicates that all of the top \( k \) samples belong to the target class, whereas a random selection yields a precision equal to the number of target samples in the validation set (\( m_2 \)) divided by the total number of samples in validation set (\( m_1 + m_2 \)). Although the Precision metric is intuitive, even if the model perfectly screens all valid samples at the top, Precision@k cannot exceed $m/k$, where $m$ is the number of target samples (true positives). When the temperature interval is very narrow, the number of true positives is small, leading to a low Precision metric. This reflects the metric’s sensitivity to data distribution.

These metrics were used to fine-tune the hyperparameters of \textit{DeepGlassNet}, including the dimensions of latent factors, embedding dimensions, the number of neurons in linear layers, learning rate, weight decay, dropout rates, activation functions, and the choice of optimizer to achieve optimal performance.

\subsubsection{Testing/Screening}

We employed a combinatorial enumeration strategy to explore the glass compositional space and obtain theoretically valid compositions for screening. Based on the fundamental design principles of oxide glasses, the composition must adhere to the following mass ratio constraints to ensure glass-forming ability and stability: the mass fraction of network formers must exceed 50 wt\%, while that of network modifiers should be maintained below 50 wt\%. The component concentration ranges and search intervals were strategically configured: \ce{SiO2}(0–80 wt\%, 5\% increment), \ce{Al2O3} (0-10 wt\%, 5\% increment), \ce{B2O3} (0–80 wt\%, 5\% increment), CaO (0–10 wt\%, 5\% increment), \ce{Na2O} (0-30 wt\%, 5\% increment),  \ce{Li2O}(0-30 wt\%, 5\% increment), MgO(0-10 wt\%, 5\% increment), BaO(0-30 wt\%, 5\% increment), and ZnO(0-30 wt\%, 5\% increment). A rigorous normalization constraint enforced total compositional mass balance ($\sum$components = 100 wt\%). This protocol generated about 39k theoretically valid glass compositions (available in \verb|test.csv|) for screening, establishing a comprehensive design space for two-phase optimization: (1) Predictive identification of compositions meeting target glass transition temperatures (T$_g$) via machine learning models, and (2) Experimental validation of top candidates. The search interval strategy balances computational efficiency with compositional resolution by assigning: broad concentration ranges to network formers (e.g., \ce{SiO2}, \ce{B2O3}), narrower concentration ranges to network modifiers/intermediates.

\par We defined the 300–400°C, 400–500°C, and 500–600°C ranges as the target T$_g$ intervals for sample screening. For each target interval, a separate model was trained. Upon obtaining the optimal DeepGlassNet model (exhibiting the highest AUC performance), it was saved and the corresponding target class centers were calculated. All test set samples were normalized (without augmentation) and input into this saved model to extract their feature representations. Within each target T$_g$ screening interval, the 10 samples closest to the class center were selected based on their component proportions for material preparation. This screening process establishes a data-driven approach for identifying optimal compositions. As these selected compositions represent novel glasses, we experimentally measured their actual glass transition temperatures (T$_g$). This dataset serves as a foundational resource for machine learning-driven material discovery and inverse design applications.

\subsubsection{Materials and Methods}
Based on the screening results, we selected two compositions from the top-10 closest to each target temperature center for preparation. The glass preparation followed these detailed steps: chemically pure starting materials all sourced
from Sinopharm Chemical Reagent Co., Ltd., China., including silicon dioxide (\ce{SiO2}), aluminum oxide (\ce{Al2O3}), boron oxide (\ce{B2O3}), sodium carbonate (\ce{Na2CO3}), lithium carbonate (\ce{Li2CO3}), zinc oxide (ZnO), strontium carbonate (\ce{BaCO3}), calcium oxide (CaO), magnesium oxide (MgO), were weighed according to the selected compositions. An appropriate amount of anhydrous ethanol was added to the raw materials, which were then placed in a QM-QX4L planetary ball mill and mixed at 400 rpm for 8 hours. The mixed powder was transferred to a high-temperature resistance furnace and melted at 1000-1550°C for 1-4 hours to ensure thorough homogenization and eliminate any compositional inhomogeneity. The molten glass was then water-quenched, and the glass samples were collected for testing. The crystalline phase and morphology of the samples were characterized using X-ray diffraction (XRD, X’Pert Pro MPD, Philips, Netherlands). Differential Scanning Calorimetry (DSC) was conducted using a NETZSCH STA449 differential scanning calorimeter, heating the samples at a rate of 10 K/min to the designated temperature. 
\section{Results}
\subsection{ML Results}
This section presents the evaluation results of \textit{DeepGlassNet} on the validation set and the insights derived from them. We first introduce the overall performance, then compare it with classical machine learning algorithms, and finally discuss the impact of the key module design of \textit{DeepGlassNet} on its performance.
\begin{figure*}[!h]
	\centering
	\vspace{-5pt}
	\subfigure[Training loss and validation AUC values across different epochs.]{\includegraphics[width=0.49\textwidth]{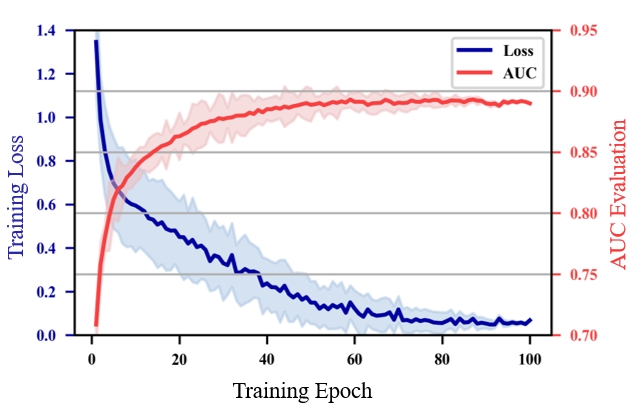}\label{fig:lossauc}}
	\subfigure[Validation ROC curves across different  epochs.]{\includegraphics[width=0.49\textwidth]{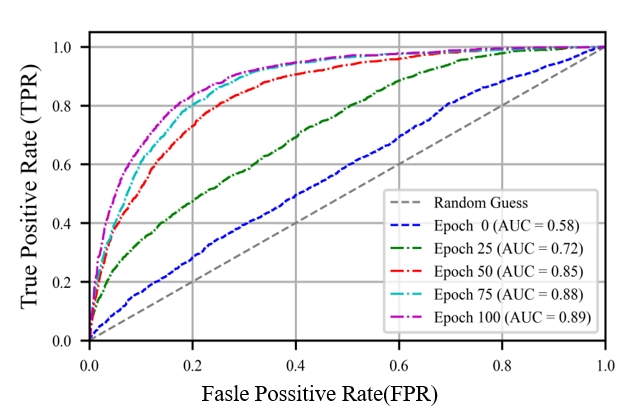}\label{fig:roc}}
	\caption{Visualization of the self-supervised learning process within the target temperature interval of 500–600°C.}
	\label{fig:vis}

\end{figure*}
\subsubsection{Overall Performance}
To comprehensively evaluate the model's performance, we established various length of the T$_g$ interval for screening, specifically 100°C, 50°C, 20°C, and 10°C. Each T$_g$ interval length includes multiple temperature ranges, as illustrated in Fig.~\ref{Fig:TW1}. The narrower the T$_g$ bandwidth, the more challenging the screening task becomes. Subsequently, samples with T$_g$ values falling within these ranges were labeled as the target class, while others were labeled as the non-target class for training and evaluation.

We then assessed the performance of \textit{DeepGlassNet} on the validation set, which comprised unseen data, using the AUC and Precision@100 metrics. The model's performance was depicted in Fig.~\ref{Fig:TW1}. This evaluation was conducted in comparison with classical algorithms, including Multilayer Perceptron (MLP), Support Vector Machine (SVM), K-Nearest Neighbors (KNN), Random Forest (RF), and Decision Tree (DT). To ensure fairness, all baseline models were rigorously optimized using the same training/validation splits as \textit{DeepGlassNet}.  Each model also underwent extensive hyperparameter tuning via grid search (see Supplementary Material for details), explicitly tailored to maximize the Area Under the Curve (AUC)—consistent with our primary evaluation criterion. The comparative results show that \textit{DeepGlassNet} significantly outperforms these classical machine learning methods across all targeted T\(_g\) intervals.

\subsubsection{Visualization of Classification Performance}

Material discovery and composition screening essentially constitute classification tasks, with the primary objective of pinpointing potential target samples. Hence, the classification proficiency of machine learning models employed in material discovery is of utmost significance. The efficacy of a model's classification capability is intuitively manifested by its capacity to position samples of the same class in close proximity to one another within the feature space. When the feature representations of target class samples exhibit a high degree of clustering, an unseen non-target sample, whose feature representation will be mapped at a greater distance from the center of the target class, will have a lower probability of being selected via top-k search, thus averting erroneous model screening.

To illustrate this, Fig.~\ref{Fig:visual} shows the TSNE visualization of the feature representations learned by \textit{DeepGlassNet} at different training stages within the target temperature range of 500–600°C. In this figure, red points represent the features of target class, while blue points represent features of non-target classes. At the initial stage of training, the feature representations of both classes are randomly distributed, showing no obvious clustering characteristics. However, as training progresses, it becomes evident that the features learned by \textit{DeepGlassNet} exhibit significant clustering features, allowing for an easy distinction between the features of target samples and those of non-target samples.

\subsubsection{Visualization of Training Process}
Fig.~\ref{fig:vis} demonstrates the visualization of the self-supervised learning process within the target temperature range of 500–600°C. Our self-supervised learning framework is meticulously designed to maximize the AUC classification metric by optimizing a contrastive loss function. To substantiate this claim, we conducted a rigorous analysis of the changes in both the AUC and loss values as the training progressed in Fig.~\ref{fig:lossauc}. The results of this analysis unveiled a clear and consistent trend: as the loss value decreased, the AUC increased steadily. This observation not only validates our design choice but also demonstrates the effectiveness of the contrastive loss in enhancing the model's classification performance. Furthermore, to provide a more intuitive and comprehensive understanding of the model's performance during training, we presented the ROC curves at different training epochs in Fig.~\ref{fig:roc}. The ROC curve, which plots the true positive rate against the false positive rate at various threshold settings, serves as a graphical representation of the model's discriminative ability. The area under the ROC curve (AUC) provides a quantitative measure of this ability. As training proceeds, it is evident from Fig.~\ref{fig:lossauc} that the AUC increases progressively, indicating that the model's ability to distinguish between positive and negative classes improves over time.

\begin{figure*}[!h]
	\centering
	\includegraphics[width=\textwidth]{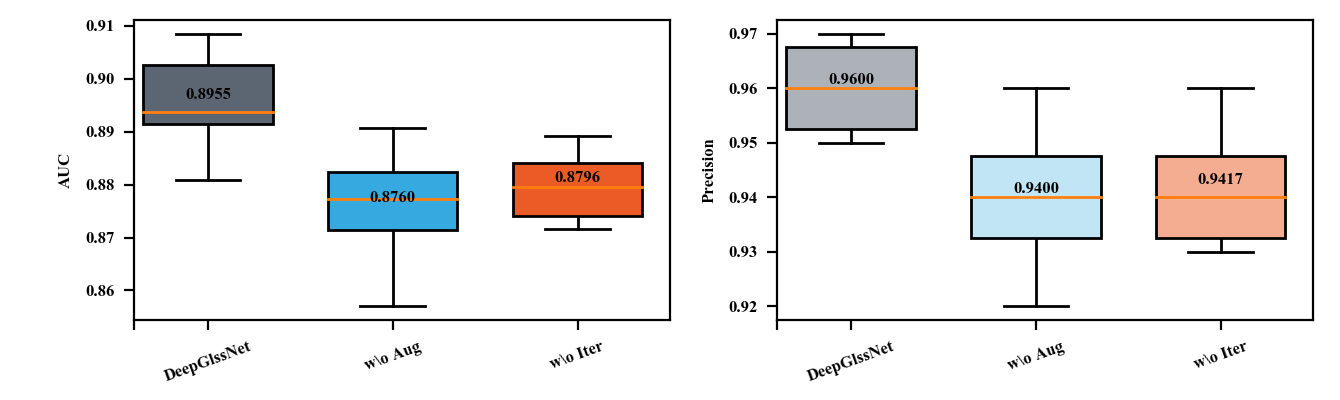}
	\caption{Results of the ablation study conducted within the temperature range of 500–600°C.}
	\label{Fig:ab}
\end{figure*}
\subsubsection{Discussions}

We tailor two key custom modules specifically for the composition screening task: (1) a graph convolution and self-attention module designed to extract features of component interactions, and (2) a data augmentation module aimed at enhancing the model's robustness against errors. In this section, we delve into the impact of these customized modules on model performance through ablation studies.

Starting with the standard \textit{DeepGlassNet}, we first removed the data augmentation module and assessed the performance of the resultant model, denoted as "\textit{DeepGlassNet} w/o Augmentation." Subsequently, we conducted an experiment by removing the interaction feature extraction module (comprising graph convolution and attention mechanisms) to eliminate the model's capability to capture high-order interactions. The resulting model, denoted as "\textit{DeepGlassNet} w/o Iter," was then evaluated.
Fig.~\ref{Fig:ab} presents ablation study results for the 500–600°C temperature range. Removal of either the interaction feature extraction module or data augmentation module significantly degrades performance relative to the complete model (p<0.05). Statistical validation through one-way ANOVA with post-hoc Tukey HSD testing confirms that both modules independently enhance model performance, while their ablation produces statistically indistinguishable degradation. This equivalence underscores their unique, non-redundant functional contributions.
Critically, our asymptotic-theory-based data augmentation introduces a novel conceptual approach to systematically model measurement noise in composition data. By modeling error propagation via the central limit theorem, it provides a physics-aware framework adaptable to other material domains—a key theoretical contribution beyond empirical tricks.


\begin{figure}[!h]
	\centering
	\includegraphics[width=0.49\textwidth]{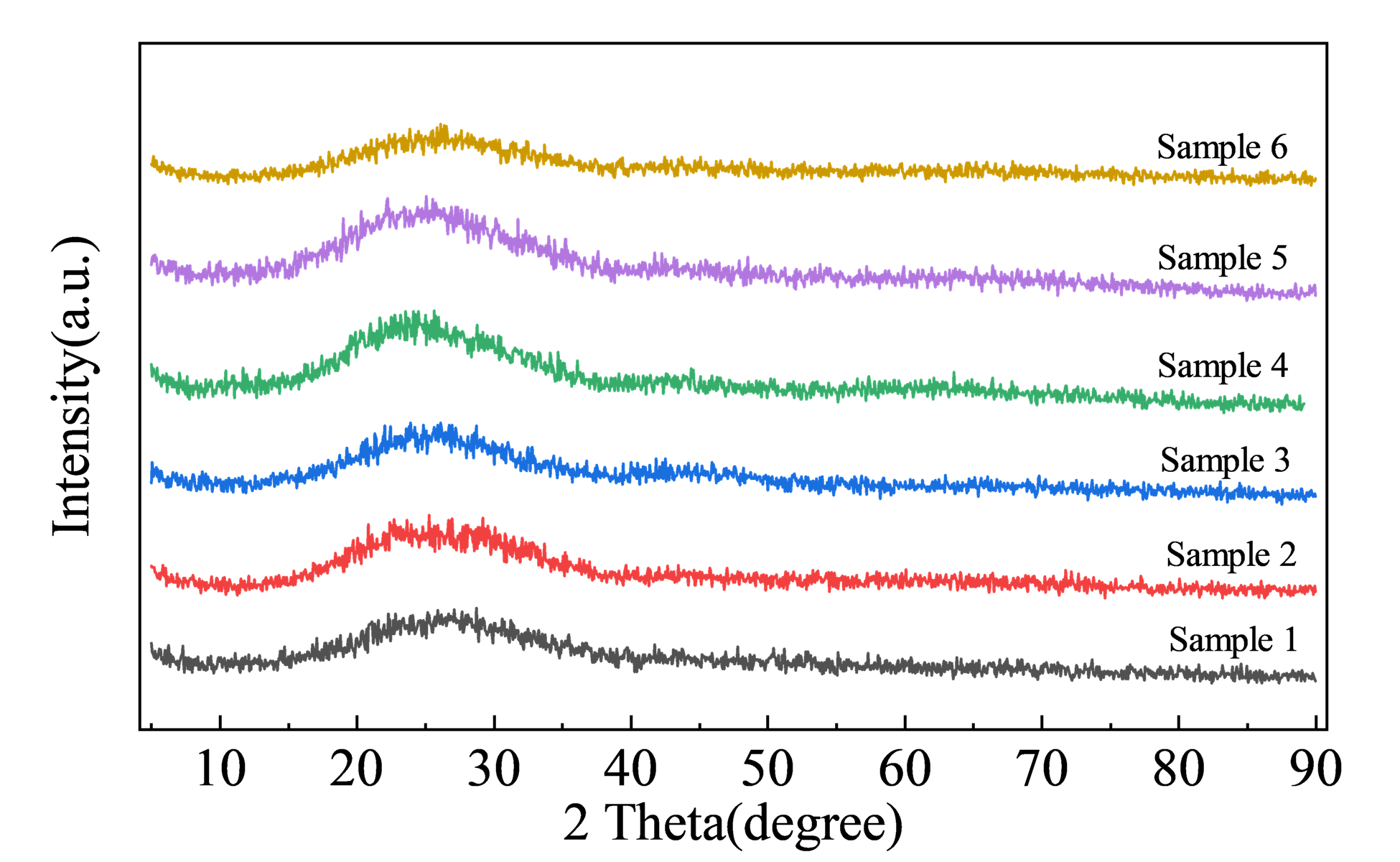}
	\caption{XRD patterns of the samples prepared by screening}
	\label{Fig:XRD}
\end{figure}
\subsection{Experimental Results and Disscussion}

\begin{figure*}[!h]
	\centering
	\includegraphics[width=0.325\textwidth]{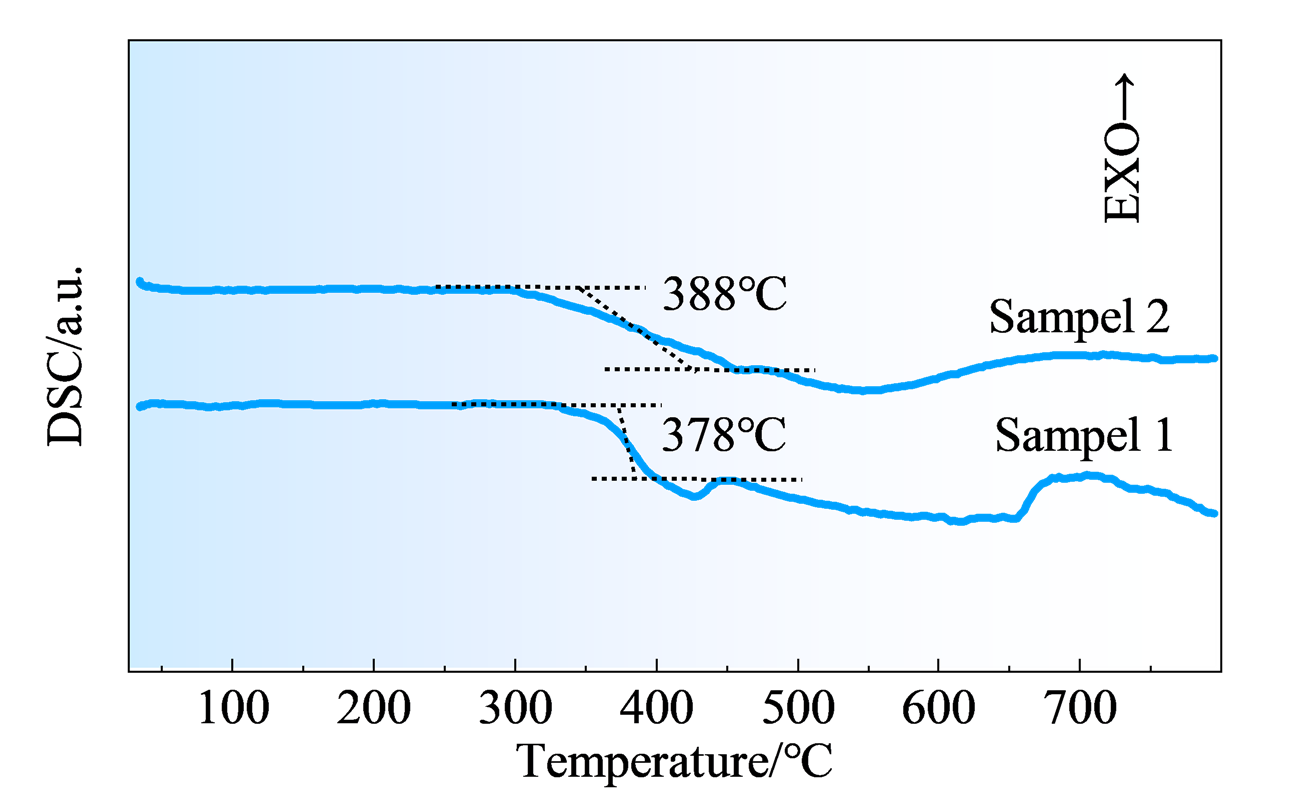}
	\includegraphics[width=0.325\textwidth]{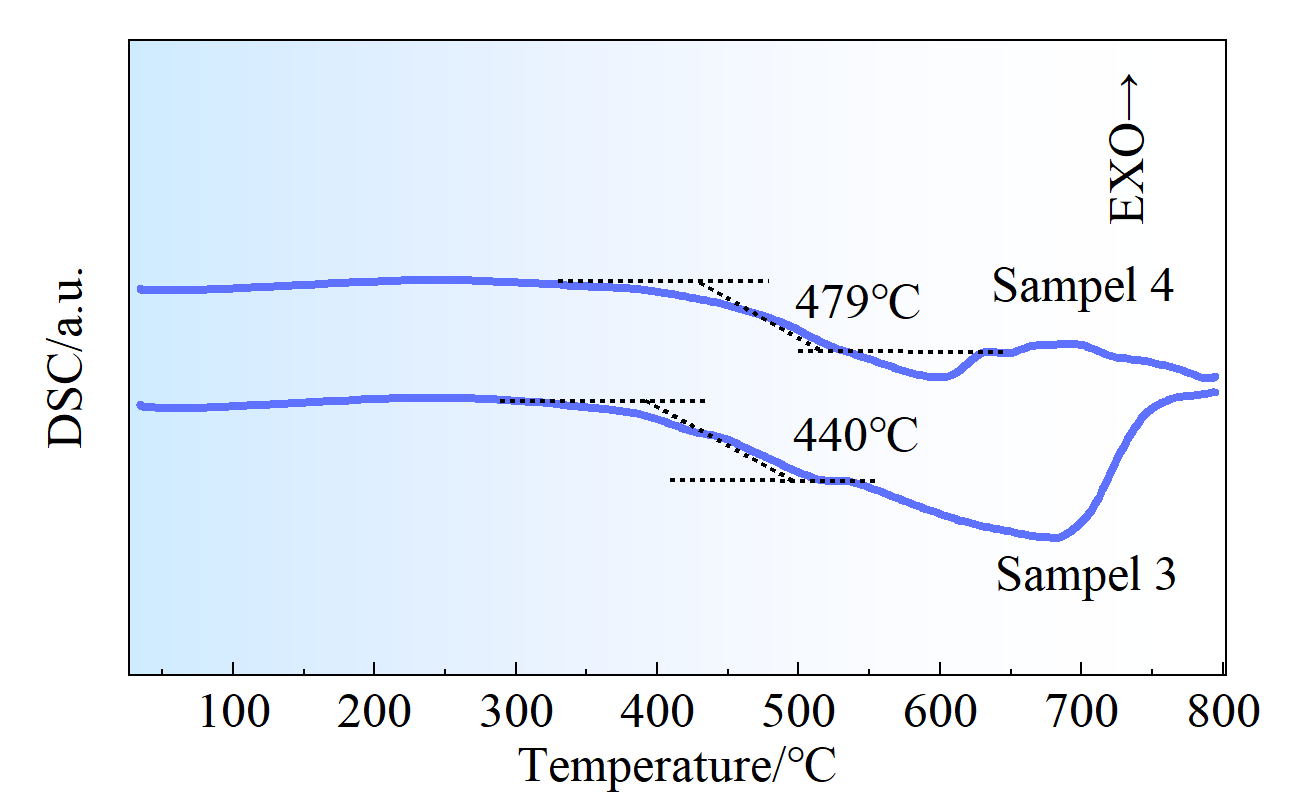}
	\includegraphics[width=0.325\textwidth]{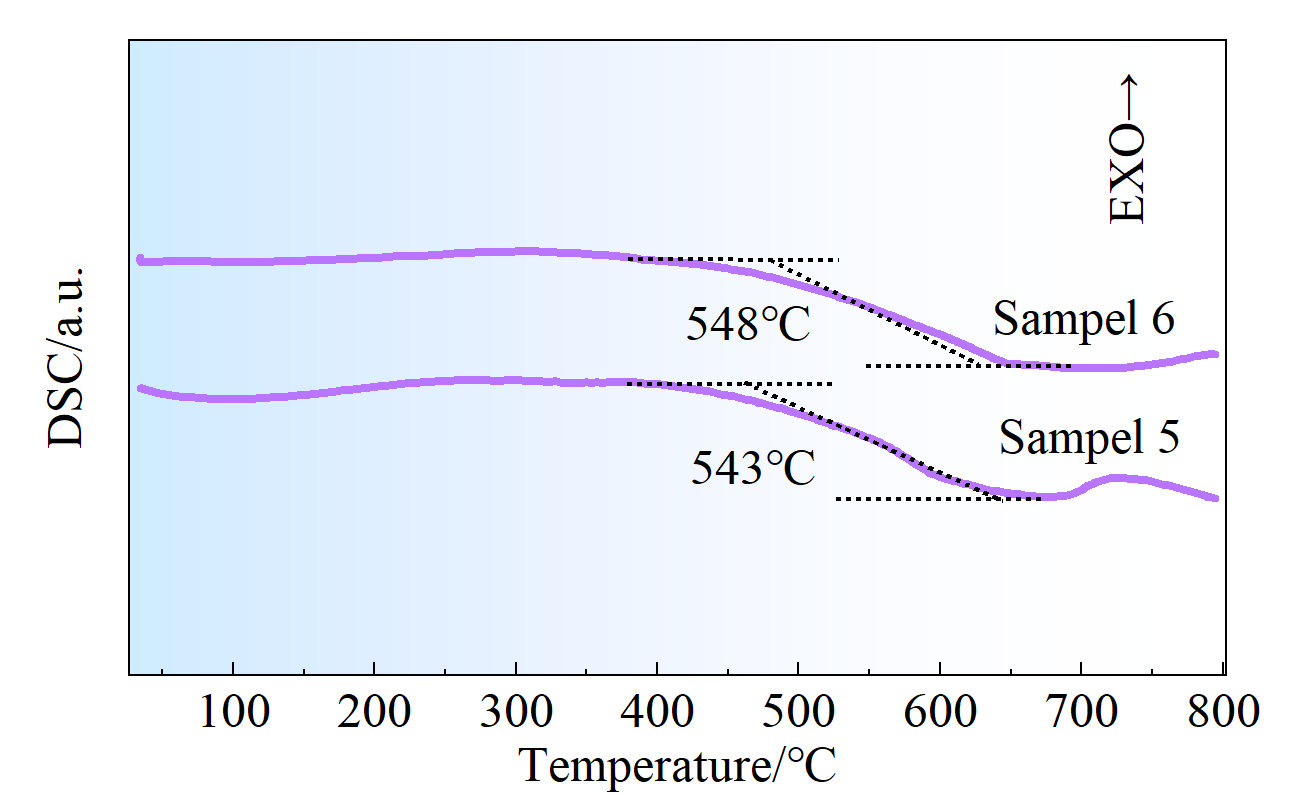}
	\caption{DSC curves of the prepared samples.}
	\label{Fig:Tg}
\end{figure*}

Samples were prepared and characterized with screening target glass transition temperature ranges of 300-400°C, 400-500°C, and 500-600°C (see Supplementary Material for details). Samples 1 and 2 were prepared targeting the 300-400°C range, Samples 3 and 4 for the 400-500°C range, and Samples 5 and 6 for the 500-600°C range. Fig.~\ref{Fig:XRD} shows the XRD patterns of the prepared glass samples. A broad diffuse scattering halo centered at approximately 20°– 40° (2$\theta$) is observed, which is characteristic of an amorphous structure lacking long-range order, thereby confirming the glassy nature of the samples. Fig.~\ref{Fig:Tg} presents the DSC curves of the screening samples at a heating rate of 10 K/min. Analysis of the DSC curves reveals that the glass transition temperatures of Sample 1 and Sample 2, which were screened for the 300-400°C range, are 378°C and 388°C respectively. Sample 1 and 2 falls within the target range. For Samples 3 and 4, prepared with the 400-500°C range as the target, their glass transition temperatures are 440°C and 479°C, both within the target range. Samples 5 and 6, targeted at the 500 -600°C range, have glass transition temperatures of 543°C and 548°C, also within the target range. 

\section{Conclusion}
\par In this study, we developed a self-supervised learning framework specifically designed to screen multicomponent glass compositions within predefined glass transition temperature (T$_g$) intervals. Our key contributions are fourfold: 
(a) Reformulating the Compositional Screening Task: We reformulate the  compositional screening task as a classification problem, aiming to predict whether the T$_g$ of a specific composition falls within a predefined range;
(b) Design of a Self-Supervised Learning Framework: We develop a self-supervised learning framework that optimizes the Area Under the Curve (AUC), with the potential for extension to other composition screening tasks;
(c) Design of Data Augmentation Method: We design a data augmentation method grounded in asymptotic theory, aiming to increase the number of training samples and enhance the model's robustness against noise;
(d) Development of \textit{DeepGlassNet}: We design \textit{DeepGlassNet}, a backbone encoder that captures the complex influences of compositional interactions on glass properties by encoding  higher-order features among different components.
Evaluation results on the validation set demonstrate that our method consistently outperforms traditional models, including K-Nearest Neighbors (KNN), Random Forest (RF), Support Vector Machine (SVM), Decision Trees (DT), and Multi-Layer Perceptron (MLP), across various T$_g$ temperature intervals. Additionally, we successfully identified samples with T$_g$ values within the ranges of 300–400°C, 400–500°C, and 500–600°C from a test set comprising 39k unseen potential composition combinations. Our framework is easily extendable to other multicomponent material screenings, providing an advanced methodology for the efficient design of multicomponent glasses and establishing a solid foundation for the application of self-supervised learning in various material discovery tasks.

\normalem
\bibliographystyle{IEEEtran}
\bibliography{refs}

\vfill
\end{document}